\documentclass[12pt]{article}
\usepackage[dvips]{graphicx}
\begin{document}
\begin{flushright}
UT-Komaba 99-17  \\
\end{flushright} 
\begin{center} 
{\Large{\bf  Lattice QCD with the Overlap
Fermions at }}\\
\vskip 0.15cm
{\Large{\bf Strong Gauge Coupling}}
\vskip 1.5cm

{\Large  Ikuo Ichinose\footnote{e-mail 
 address: ikuo@hep1.c.u-tokyo.ac.jp}and 
 Keiichi Nagao\footnote{e-mail
 address: nagao@hep1.c.u-tokyo.ac.jp}}  
\vskip 0.5cm
 
Institute of Physics, University of Tokyo, Komaba,
  Tokyo, 153-8902 Japan 
\end{center}
\vskip 3cm
\begin{center} 
\begin{bf}
Abstract
\end{bf}
\end{center}
We generalize overlap fermion by Narayanan and Neuberger
by introducing a hopping parameter $t$.
This lattice fermion has desirable properties as the original
overlap fermion.
We expand ``Dirac" operator of this fermion in powers of $t$.
Higher-order terms of $t$ are long-distance terms and this
$t$-expansion is a kind of the hopping expansion.
It is shown that the Ginsparg-Wilson relation is satisfied 
at each order of $t$.
We show that this $t$-expansion is useful for study of the strong-coupling  
gauge theory.
We apply this formalism to the lattice QCD and study its chiral phase
structure at strong coupling.
We find that there are (at least) two phases one of which has
desired chiral properties of QCD.
Possible phase structure of the lattice QCD with the overlap
fermions is proposed.

\newpage
\setcounter{footnote}{0}
\section{Introduction}
Recently a very promising formulation named overlap fermion
was proposed by Narayanan and Neuberger \cite{NN,Ne} and it has been studied 
intensively.
However almost all (analytical) studies on the overlap fermion
employ the weak-coupling expansion.
Study on the overlap fermion interacting with the strong-coupling gauge 
field is desired.
Especially to clarify its phase structure is very important, e.g.,
for numerical studies and in order to take the continuum limit.
 
We shall give a way for study of strong-coupling gauge theory
of the overlap fermion.
To this end, we slightly generalize the original overlap fermion 
by introducing a hopping parameter $t$.
This lattice fermion, which we call generalized overlap (GO) fermion,
has desirable properties as the original overlap fermion.
We can expand ``Dirac operator" of the GO fermion in powers of $t$.
Higher-order terms are long-distance terms and therefore the $t$-expansion
is a kind of the hopping expansion.
The Ginsparg-Wilson (GW) relation \cite{GW} is satisfied at each order of $t$.
Strong-coupling studies of the gauge field can be applied for 
this $t$-expanded Dirac operator of the GO fermion straightforwardly.
We expect that the $t$-expansion is justified in the strong-coupling
region of gauge theory because hopping of a single quark is suppressed
by large fluctuation of the gauge field.
Moreover, the $t$-expansion reveals properties of L\"uscher's 
extended chiral symmetry which play an important role for study of
phase structure.
We then study lattice QCD with the GO fermions by using the $t$-expansion
at the strong-coupling limit.
It is expected that the strong-coupling studies give qualitatively
correct picture for the strong-coupling gauge theory like QCD.
We find that there are (at least) two phases.
One of them has desired properties of QCD and in this phase
the extended chiral symmetry can be considered as a properly
generalized chiral symmetry at finite lattice spacing.
On the other hand, the other phase has rather anomalous properties
concerning the chiral symmetry.

\section{GO fermion and $t$-expansion}

We consider $d$-dimensional square lattice with the lattice 
spacing $a$, which will be often set unity in later discussion.
Fermion variables $\bar{\psi}(n)$ and $\psi(n)$ are defined on 
site $n$ and U(N) or SU(N) gauge field $U_{\mu}(n)$ ($\mu$ is the direction
index, $\mu=1\sim d$) are defined on link $(n,\mu)$.
The GO fermion is given by the following action 
\begin{equation}
S_F=a^d\sum_{n,m}\bar{\psi}(m)D(m,n)\psi(n),
\label{SF}
\end{equation}
where the covariant derivative $D(m,n)$ is defined as
\begin{eqnarray}
D&=&{1\over a}\Big(1+X{1 \over \sqrt{X^{\dagger}X}}\Big),  \nonumber  \\
X_{mn}&=&\gamma_{\mu}C_{\mu}(t;m,n)+B(t;m,n),  \nonumber   \\
C_{\mu}(t;m,n)&=&{t \over 2a}\Big[\delta_{m+\mu,n}U_{\mu}(m)-
\delta_{m,n+\mu}U^{\dagger}_{\mu}(n)\Big],  \nonumber  \\
B(t;m,n)&=&-{M_0\over a}+{r\over 2a}\sum_{\mu}\Big[2\delta_{n,m}
-t \delta_{m+\mu,n}U_{\mu}(m)-t \delta_{m,n+\mu}U^{\dagger}_{\mu}(n)\Big],
\label{covD}
\end{eqnarray} 
where $r$ and $M_0$ are 
dimensionless nonvanishing free parameters of the overlap lattice fermion 
formalism.
We have introduced a new parameter $t$.
The original overlap fermion corresponds to $t=1$.
For notational simplicity, we define 
\begin{equation}
A\equiv {1\over a}(dr-M_0), \;\; 
B\equiv {rt \over 2a},  \;\;
C\equiv {t\over 2a}.
\label{ABC}
\end{equation}
It is verified that propagator at tree level $U_{\mu}(n)=1$
has no species doublers for the parameter region 
$(1-t)dr<M_0<(d-dt+2t)r$.
This parameter region is renormalized by the interactions.
Therefore it is important to study phase structure of the system
in wide parameter region of $M_0$ and the gauge coupling 
constant $g^2$ with fixed values of $r$ and $t$, for example.
It is also verified by the weak-coupling expansion that the GO fermion
generates the ordinary chiral anomaly as it is desired.
Actually it is verified that the $t$-dependence of $D(m,n)$ is absorbed
by redefinition of the parameter $M_0$.
In this sense the GO fermion is not quite new.

In this paper we shall expand the GO fermion operator (\ref{covD})
in powers of $t$ assuming that $t$ is small.
As we shall show, this $t$-expansion is a kind of the hopping expansion
and then we expect that the $t$-expansion is justified and suitable
for the strong-coupling gauge theory.
At strong-coupling region, movement of a single quark is suppressed
by the strong fluctuation of the gauge field, i.e., 
$\langle U_\mu(m)\rangle \sim 0$, 
and the number 
of paths in the random-walk representation of correlation function
of gauge-invariant composite fields is much smaller than that 
of weak-coupling cases.\footnote{However, we also expect that the 
$t$-expansion has a finite convergence radius for smooth configurations
of fields, for higher-order terms contain higher-powers of the difference
operator $\Gamma^-_\mu(m,n)$ defined by Eq.(\ref{-+}).
In the strong-coupling phase, on the other hand, the good convergence 
of the $t$-expansion is expected {\em after} integration over the fluctuating
gauge field.} 

For notational simplicity, let us define the following
quantities,
\begin{eqnarray}
\Gamma^-_\mu (m,n)&=&\delta_{m+\mu,n}U_{\mu}(m)-
\delta_{m,n+\mu}U^{\dagger}_{\mu}(n),  \nonumber  \\
\Gamma^+_\mu(m,n)&=&\delta_{m+\mu,n}U_{\mu}(m)+
\delta_{m,n+\mu}U^{\dagger}_{\mu}(n).
\label{-+}
\end{eqnarray}
In terms of the above quantities,
\begin{equation}
X_{mn}=A\delta_{mn}+C\sum\gamma_\mu \Gamma^-_\mu (m,n)-
B\sum\Gamma^+_\mu (m,n),
\label{X2}
\end{equation}
\begin{equation}
(X^\dagger)_{mn}=A\delta_{mn}-C\sum\gamma_\mu \Gamma^-_\mu (m,n)-
B\sum\Gamma^+_\mu (m,n).
\label{X3}
\end{equation}
From Eq.(\ref{ABC}), $B,C =O(t)$ and we consider $A=O(1)$ in later
discussion.
Then it is rather straightforward to expand $D(m,n)$ in powers of
$t$,
\begin{eqnarray}
\Big(\sqrt{X^\dagger X}\Big)_{mn}&=&|A|\delta_{mn}
-{|A|B \over A}\sum\Gamma^+_{\mu}(m,n)-{C^2 \over 2|A|}
\sum\gamma_\mu\gamma_\nu \Gamma^-_\mu(m,l)\Gamma^-_\nu(l,n) \nonumber  \\
&& \;\; -{BC \over 2|A|}\sum \gamma_\mu\Big(\Gamma^+_\nu(m,l)
\Gamma^-_\mu(l,n)-\Gamma^-_\mu(m,l)\Gamma^+_\nu(l,n)\Big) 
+O(t^3), \nonumber  \\
\Bigg(X{1 \over \sqrt{X^\dagger X}}\Bigg)_{mn}&=&\mbox{sgn}(A)\delta_{mn}
+{C \over |A|}\sum \gamma_\mu\Gamma^-_\mu(m,n) \nonumber  \\
&&\;\; +{BC \over 2A|A|}\sum \gamma_\mu\Big(\Gamma^-_\mu(m,l)
   \Gamma^+_\nu(l,n) +\Gamma^+_\nu(m,l)\Gamma^-_\mu(l,n)\Big)  \nonumber   \\
&& \;\; +{C^2 \over 2A|A|}\sum\gamma_\mu\gamma_\nu \Gamma^-_\mu(m,l)  
   \Gamma^-_\nu(l,n)+O(t^3),  \nonumber   \\
aD(m,n)&=&2\theta(A)\delta_{mn}
+{C \over |A|}\sum \gamma_\mu\Gamma^-_\mu(m,n)  \nonumber   \\
&& \;\;
   +{BC \over 2A|A|}\sum \gamma_\mu\Big(\Gamma^-_\mu(m,l)
   \Gamma^+_\nu(l,n)+\Gamma^+_\nu(m,l)\Gamma^-_\mu(l,n)\Big)  \nonumber   \\
&& \;\; +{C^2 \over 2A|A|}\sum\gamma_\mu\gamma_\nu \Gamma^-_\mu(m,l)  
   \Gamma^-_\nu(l,n)+O(t^3).  
\label{tD}
\end{eqnarray}   
It is verfied that the Ginsparg-Wilson (GW) relation \cite{GW}
\begin{equation}
D\gamma_5+\gamma_5D=aD\gamma_5D,
\label{GW}
\end{equation}
is satisfied by the $t$-expanded $D(m,n)$ in Eq.(\ref{tD}) at each order of
$t$.

Action of the fermion $S_F$ in Eq.(\ref{SF}) is invariant under 
the following extended chiral transformation by L\"uscher \cite{Lus},
\begin{equation}
\delta\psi(m)=\epsilon\gamma_5\Big(\delta_{nm}-aD(m,n)\Big)\psi(n),\;\;
\delta\bar{\psi}(m)=\epsilon\bar{\psi}(m)\gamma_5,
\label{extended}
\end{equation}
where $\epsilon$ is an infinitesimal transformation parameter.
Then in the overlap fermion formalism, it is natural to
think that the extended chiral symmetry (\ref{extended})
is more fundamental than the usual chiral symmetry which is broken
at finite lattice spacing.

In terms of the $t$-expansion, the transformation (\ref{extended}) 
is given as,
\begin{eqnarray}
\delta\psi(m)&=&\epsilon\gamma_5\Big(-\mbox{sgn}(A)\delta_{mn}
-{C \over |A|}
\sum\gamma_\mu\Gamma^-_\mu(m,n)+\cdot\cdot\cdot\Big)\psi(n), \nonumber  \\
\delta\bar{\psi}(m)&=&\epsilon\bar{\psi}(m)\gamma_5.
\label{extended2}
\end{eqnarray}
This expression reveals the fact that the extended chiral 
transformation (\ref{extended})
has quite different meanings depending on the sign of the parameter $A$.
Especially for positive $A$, $\psi(m)$ and $\bar{\psi}(m)$
have the same ``extended-chiral charge" at the leading order of $t$.
This explains the appearance of the term 
$\theta(A)\sum\bar{\psi}(m)\psi(m)$ in $S_F$.
Therefore we have to discuss chiral properties of the QCD with
the GO fermions for positive and negative $A$ separately.

\section{Strong-coupling limit}

In the previous section, we $t$-expanded the ``Dirac" operator
of the GO fermion.
We show that the $t$-expansion is suitable for the strong-coupling 
studies of the lattice QCD whose action is given by,
\begin{eqnarray}
S_{tot}&=&S_G+S_{F,M},  \nonumber   \\
S_G&=&-{1\over g^2}\sum_{pl}\mbox{Tr}(UUU^\dagger U^\dagger),  \nonumber  \\
S_{F,M}&=& S_F-M_B\sum\bar{\psi}(m)\psi(m),
\label{Stot}
\end{eqnarray}
where we have added the bare mass term of quarks.
In this section, we shall study chiral structure of the above
system in the strong-coupling limit $g^2N\rightarrow \infty$,
though the strong-coupling expansion can be performed systematically.
As explained above, we must consider the cases of positive and negative 
values of $A$ separately.
In this section we mostly set the lattice spacing $a=1$.

\subsection{Negative $A$}

We shall consider U(N) gauge theory for definiteness.
The partition function of the system is given by
\begin{equation}
Z[J]=\int D\bar{\psi}D\psi DU \exp \Big\{-S_{tot}+\sum J(n)\hat{m}(n)\Big\},
\label{partition}
\end{equation}
where $[DU]$ is the Haar measure and 
\begin{eqnarray}
&& J(n)\hat{m}(n)=J^\alpha_\beta(n)\hat{m}_\alpha^\beta(n)  \nonumber  \\
&&  \hat{m}_\alpha^\beta(n)=
{1\over N}\sum_a\psi_{a,\alpha}(n)
\bar{\psi}^{a,\beta}(n),
\label{Jm}
\end{eqnarray}
with color index $a$ and spinor-flavor indices $\alpha$ and $\beta$.
The effective action $S_{eff}({\cal M})$ is defined as 
\begin{equation}
Z[J]=\int D{\cal M} e^{-S_{eff}({\cal M})+J{\cal M}}
\label{Seff}
\end{equation}
where integral over color-singlet ``meson" field ${\cal M}^\alpha_\beta$
will be defined later on.

To obtain $S_{eff}({\cal M})$, it is useful to notice that
the following combination is invariant under the extended
chiral transformation (\ref{extended2}),
\begin{eqnarray}
&& q\equiv \Big(1-{a\over 2}D\Big)\psi,\; \; \bar{q}=\bar{\psi}, 
\; \; \delta q(n)=\epsilon \gamma_5 q(n), \;\;
\delta \bar{q}(n)=\epsilon \bar{q}(n)\gamma_5,   \nonumber  \\
&&\bar{q}q(m)\bar{q}q(n)-\bar{q}\gamma_5q(m)\bar{q}\gamma_5q(n) \nonumber  \\
&& \;\; =\bar{\psi}\psi(m)\bar{\psi}\psi(n)-\bar{\psi}\gamma_5\psi(m)
\bar{\psi}\gamma_5\psi(n)  \nonumber   \\
&& \;\;+ {C\over 2|A|}\Big\{\bar{\psi}(m)\gamma_5\sum\gamma_\mu\Gamma^-_\mu
(m,l)\psi(l)\bar{\psi}\gamma_5\psi(n)  
 -\bar{\psi}(m)\sum\gamma_\mu\Gamma^-_\mu
(m,l)\psi(l)\bar{\psi}\psi(n)   \nonumber   \\
&& \; \; +(m \leftrightarrow n)\Big\}+O(t^2).
\label{inv}
\end{eqnarray}
The first step for the effective action is the one-link integral
of the gauge field,
\begin{equation}
e^{W(\bar{D},D)}=\int dU_\mu \exp\Big[\mbox{Tr}(\bar{D}_\mu U_\mu
+U^\dagger_\mu D_\mu)\Big].
\label{onelink}
\end{equation}
For the U(1) gauge group, the above integral is easily
performed as $W(\bar{D},D)=\bar{D}D-{1\over 4}(\bar{D}D)^2
+\cdot\cdot\cdot$.
For the U(N) gauge theory
$W(\bar{D},D)$ was calculated by Brezin and Gross for large $N$.
There are two ``phases" in the above one-link intergal,
and in the strong-coupling regime, which is relevant for the present study,
$W(\bar{D},D)$ is given by the following formula \cite{BG},
\begin{eqnarray}
W(\bar{D},D)&=&N^2\Big\{ -{3\over 4} -c +{2 \over N}
\sum_a (c+x_a)^{1/2}  \nonumber  \\
&& \; \; -{1\over 2N^2}\sum_{a,b}\log (
(c+x_a)^{1/2}+(c+x_b)^{1/2})\Big\},
\label{strongW}
\end{eqnarray}
where $x_a$'s are eigenvalues of ${1 \over N^2}\bar{D}D$
and a constant $c$ is implicitly given by 
\begin{equation}
1={1 \over 2N}\sum_a(c+x_a)^{-1/2}.
\label{c}
\end{equation}
In the present study, the ``Dirac operator" $D(m,n)$ is given by
(\ref{tD}), and therefore for the expansion in powers of $t$,
we set
\begin{eqnarray}
&& D^a_{\mu b}=A^a_{\mu b}+C^a_{\mu b}, \;\; 
   \bar{D}^a_{\mu b}=\bar{A}^a_{\mu b}+\bar{C}^a_{\mu b}, \nonumber  \\
&& A^a_{\mu b}={C\over |A|}\bar{\psi}_b(n+\mu)\gamma_\mu \psi^a(n), 
\nonumber  \\
&& \bar{A}^a_{\mu b}=-{C\over |A|}\bar{\psi}_b(n)
\gamma_\mu \psi^a(n+\mu), 
\label{D}
\end{eqnarray}
where $C_\mu$ and $\bar{C}_\mu$ are sources.
The following identity is useful which is proved for an arbitrary  
regular function $f(x)$,
\begin{eqnarray}
{1\over N}\sum_a f(x_a)&=&f(0)-{1\over N}\mbox{Tr}
[f(-\lambda)-f(0)]  \nonumber   \\
&&-{1\over N^3}\Big({C\over |A|}\Big)\bar{C}_\mu(n)^b_a
\mbox{Tr}\Big[\Big(\psi^a(n)
\bar{\psi}_b(n+\mu)\Big)\gamma_\mu f'(-\lambda)\Big] \nonumber  \\
&& +{1\over N^3}\Big({C\over |A|}\Big)C_\mu(n)^b_a
\mbox{Tr}\Big[\Big(\psi^a(n+\mu)
\bar{\psi}_b(n)\Big)\gamma_\mu f'(-\lambda') \Big] \nonumber  \\
&&+O(C^2_\mu),
\label{identity}
\end{eqnarray}
where 
\begin{eqnarray}
\lambda&=&\lambda_\mu (n)=\Big({C\over A}\Big)^2\hat{m}(n)\gamma_\mu
\hat{m}(n+\mu)\gamma_\mu,\nonumber \\ 
\lambda'&=&\lambda'_\mu (n)=\Big({C\over A}\Big)^2\hat{m}(n+\mu)
\gamma_\mu \hat{m}(n) \gamma_\mu.
\label{lambda}
\end{eqnarray}
From Eqs.(\ref{strongW}) and (\ref{identity}) we obtain the one-link 
integral as follows \cite{I,KS},
\begin{eqnarray}
{1 \over N^2}W(\bar{D},D) &=& -{1\over N} \mbox{Tr}[(1-4\lambda)^{1/2}
-1]+{1\over N}\mbox{Tr}[\log{\frac{1+(1-4\lambda)^{1/2}}{2}}]  \nonumber  \\
&&-{2\over N^3}\Big({C\over |A|}\Big)\bar{C}_\mu(n)^b_a
\mbox{Tr}\Big[\Big(\psi^a(n)
\bar{\psi}_b(n+\mu)\Big) \gamma_\mu (1+(1-4\lambda)^{1/2})^{-1}
\Big] \nonumber  \\
&&+{2\over N^3}\Big({C\over |A|}\Big)C_\mu(n)^b_a
\mbox{Tr}\Big[\Big(\psi^a(n+\mu)
\bar{\psi}_b(n)\Big) \gamma_\mu (1+(1-4\lambda')^{1/2})^{-1} 
 \Big] \nonumber  \\
&&+O(C_\mu^2).
\label{WDD}
\end{eqnarray}
From (\ref{WDD}), the expectation value of the gauge field is given
as follows in the strong-coupling limit,
\begin{eqnarray}
\langle U^a_{\mu b}(n) \rangle_U &=&-{2 \over N}
\Big({C\over |A|}\Big)
\mbox{Tr}\Big[\Big(\psi^a(n)
\bar{\psi}_b(n+\mu)\Big)\gamma_\mu (1+(1-4\lambda)^{1/2})^{-1}
\Big] \nonumber  \\
\langle U^{\dagger a}_{\mu b}(n) \rangle_U &=& {2 \over N}
\Big({C\over |A|}\Big)\mbox{Tr}\Big[\Big(\psi^a(n+\mu)
\bar{\psi}_b(n)\Big) \gamma_\mu (1+(1-4\lambda')^{1/2})^{-1}  \Big],
\label{expectU}
\end{eqnarray}
where $\langle \cdot\cdot\cdot \rangle_U$ denotes 
average over the gauge field $U_\mu(n)$.

In Ref.\cite{chand} spontaneous
symmetry breakdown of the extended chiral symmetry is argued
and an order parameter is given by
\begin{equation}
\langle \bar{q}q \rangle=\langle \bar{\psi}(1-{a \over 2}D)\psi\rangle.
\label{OP}
\end{equation}
Nambu-Goldstone bosons appear in the channel 
$(\bar{\psi}\gamma_5\psi)$ and their mass $m^2_\pi \propto M_B$
(in the flavor-nonsinglet channel).\footnote{Two composite fields
$(\bar{\psi}\gamma_5\psi)$ and $(\bar{q}\gamma_5q)$ differ with
each other only invariant quantity under the extended chiral 
transformation.
Therefore $(\bar{q}\gamma_5q)$ can be also considered as 
Nambu-Goldstone bosons.}
The above criterion is verified by using solvable models \cite{IN2}.

We shall calculate the order parameter (\ref{OP}) in the present
formalism in the strong-coupling limit.
After the integral over the gauge field, the partition function
is given by the functional integral over the fermions with a new action
which is a functional of the color-singlet composite mesons
$\hat{m}^\alpha_\beta(n)$.
Moreover because of the extended chiral symmetry, the action of 
$\hat{m}^\alpha_\beta(n)$
depends on the chiral invariants like (\ref{inv}) with the replacement
of gauge fields as in (\ref{expectU}).
Flavor-singlet extended chiral symmetry is explicitly broken by the measure
of the fermion path integral.
Effects of anomaly will appear in the next-leading
order of $1/N$ and the noninvariance of the fermion measure is
related with the U(1) problem \cite{Lus,IN-QCD}.

Elementary meson fields are introduced through the identity
like (up to irrelevant constants),
\begin{eqnarray}
\int d\bar{\psi}d\psi \exp\Big({1\over N}J^\beta_\alpha\psi^\alpha_a
\bar{\psi}^a_\beta\Big) &=& \Big(\mbox{det}J\Big)^N  \nonumber   \\
&=&\oint d{\cal M}
\Big(\mbox{det}{\cal M} \Big)^{-N}\cdot e^{J\cdot{\cal M}},
\label{int-meson}
\end{eqnarray}
where the integral over ${\cal M}$ is defined by the contour integral,
i.e., ${\cal M}$ is polar-decomposed as ${\cal M}=RV$ with positive-definite
Hermitian matrix $R$ and unitary matrix $V$, and 
$\oint d{\cal M}\equiv \int dV$ with the Haar measure of U($N_{sf}$)
($N_{sf}$ is the dimension of the spinor-flavor index) \cite{KS}.
From (\ref{int-meson}), there appear additional terms like
$(N\mbox{Tr}\log {\cal M})$ in the effective action.
Detailed study of the low-energy effective theory of hadrons
will be given in a forthcoming paper \cite{IN-QCD}.
In this paper we shall calculate the order parameter (\ref{OP}).
From the discussion of the extended chiral symmetry given above and in
Ref.\cite{chand}, it is obvious that Nambu-Goldstone pions
appear if the spontaneous chiral symmetry breakdown
$\langle \bar{\psi}(1-{a \over 2}D)\psi\rangle\neq 0$ occures.

We assume the pattern of symmetry breaking for simplicity.
By the existence of the bare mass term of quarks, 
\begin{equation}
\langle {\cal M}^\alpha_\beta\rangle=v \delta^\alpha_\beta,
\label{VEV}
\end{equation}
where $v$ is some constant which will be calculated from now.
From (\ref{lambda}) and (\ref{VEV}),
\begin{equation}
\lambda=\lambda'=\Big({C\over A}\Big)^2v^2.
\end{equation}
Then effective potential of $v$ or $\lambda$ is obtained as
\begin{eqnarray}
{V_{eff}(\langle\bar{\psi}\psi\rangle) \over NN_{sf}}&=&
{1\over 2}\log (4\lambda)+d\Big\{(1-4\lambda)^{1/2}
-\log[1+(1-4\lambda)^{1/2}] \Big\}  \nonumber  \\
&&\;\; +M_Bv+ O(t).
\label{effV}
\end{eqnarray}
From Eq.(\ref{effV}), we obtain for vanishing quark mass
\begin{equation}
\lambda={2d-1 \over 4d^2}+O(t), \;\; 
v={|A| \over 2C}\sqrt{{2d-1 \over d^2}} +O(t^0).
\label{VEV2}
\end{equation}
Higher-order terms of $t$ can be systematically calculated
in the present formalism \cite{IN-QCD}.
From Eq.(\ref{VEV2}), we have  
 \begin{equation}
\langle \bar{\psi}(1-{a \over 2}D)\psi\rangle =NN_{sf}
{|A| \over 2C}\sqrt{{2d-1 \over d^2}} +O(t^0),
\end{equation}
and therefore spontaneous symmetry breaking of the extended chiral 
symmetry occures at strong coupling.


\section{Higher-order terms}
 
In the previous section, we calculated chiral condensate 
$\langle\bar{\psi}\psi\rangle$ at the leading order of $t$.
In this section we shall consider higher-order terms of the 
effective potential.
Especially we show that contributions from the terms 
in the action, which are determined by the GW relation (\ref{GW})
from the lower-order terms,
can be summed up.
Example of such a term is 
\begin{equation}
{C^2 \over 2A|A|}\sum\bar{\psi}(m)\gamma_\mu\gamma_\nu \Gamma^-_\mu(m,l)
\Gamma^-_\nu(l,n)\psi(n).
\label{gamma2}
\end{equation}
in the action $S_F$.
This term is completely determined by the GW relation from the term
$$
{C \over |A|}\sum \bar{\psi}(m)\gamma_\mu\Gamma^-_\mu(m,n)\psi(n)
$$
in the action. 

It is tedious but straightforward to verify that contribution from 
the term (\ref{gamma2}) changes the effective potential in (\ref{effV})
as
\begin{equation}
V_{eff}(\langle\bar{\psi}\psi\rangle) \Rightarrow 
V_{eff}(\langle\bar{q}q\rangle).
\label{effVq}
\end{equation}
To this end we use equations like 
\begin{eqnarray}
\langle U^a_{\mu b}(m)U^b_{\nu c}(m+\mu)\rangle_U &=&
\Big({C\over NA}\Big)^2\mbox{Tr}[\psi^a(m)\bar{\psi}_b(m+\mu)
\gamma_\mu g'(\lambda_\mu(m))]  \nonumber   \\
&& \;\times
\mbox{Tr}[\psi^b(m+\mu)\bar{\psi}_c(m+\mu+\nu)
\gamma_\nu g'(\lambda_\nu(m+\mu))]  ,
\end{eqnarray}
where 
\begin{eqnarray} 
g(x)&=&-(1-4x)^{1/2}+1+\log[\frac{1+(1-4x)^{1/2}}{2}], \nonumber  \\
g'(x)&=& 2(1+(1-4x)^{1/2})^{-1}.
\end{eqnarray}
Then the term (\ref{gamma2}) generate terms like\footnote{Here
we have neglected terms proportional to $(\bar{\psi}\gamma_5
\psi)(m)$.This is vanishing in the condensation pattern (\ref{VEV}) 
which we assume in the rest of this section.}
\begin{eqnarray}
&& {1 \over 2N} \Bigg({C\over A}\Bigg)^4\hat{m}(m)
  \Big(\bar{\psi}_a(m+\mu)\gamma_\nu\psi^b(m+\mu+\nu)\Big)  \nonumber  \\
&& \;\; \times \Big(\bar{\psi}_b(m+\mu+\nu)\gamma_\nu\psi^a(m+\mu)\Big) 
g'(\lambda_\mu(m))g'(\lambda_\nu(m+\mu))+\cdot\cdot\cdot.
\label{t-High1}
\end{eqnarray}
On the other hand, 
\begin{equation}
\hat{m}(m)
\Big(\bar{q}q(m+\mu)-\bar{\psi}\psi(m+\mu)\Big)=-\hat{m}(m)
\Big({C\over 2|A|}\sum
\bar{\psi}(m+\mu)\gamma_\nu\Gamma^-_\nu(m+\mu,n)\psi(n)+\cdots\Big).
\end{equation}
After the path-integral over the gauge fields,
\begin{eqnarray}
\hat{m}(m)
\Big(\bar{q}q(m+\mu)-\bar{\psi}\psi(m+\mu)\Big)&\Rightarrow&-{1\over 2N}
\Big({C\over A}\Big)^2\hat{m}(m)
  \Big(\bar{\psi}_a(m+\mu)\gamma_\nu\psi^b(m+\mu+\nu)\Big)  \nonumber  \\
&& \; \times \Big(\bar{\psi}_b(m+\mu+\nu)\gamma_\nu\psi^a(m+\mu)\Big) 
g'(\lambda_\nu(m+\mu))  \nonumber  \\
&& \;  +\cdot\cdot\cdot.
\label{t-High2}
\end{eqnarray}
Then from Eqs.(\ref{lambda}), (\ref{t-High1}) and (\ref{t-High2}),
one can see that the contribution from the high-order term (\ref{gamma2})
simply replaces $\lambda_\mu(m)$ in the effective potential with
$$
\Big({C \over A}\Big)^2q\bar{q}(m+\mu)q\bar{q}(m).
$$
We have verified this result only at the lowest-nontrivial order, but we 
expect that it is correct at all orders of $t$.

For completeness, we have to examine the term which comes from
the ${\cal M}$-integral and contributes to the effective potential.
We evaluate the following integral instead of Eq.(\ref{int-meson}),
\begin{equation}
\int d\bar{\psi}d\psi \exp\Big({1\over N}Jq\bar{q}\Big).
\label{int-meson2}
\end{equation}
We can change the measure of the above path-integral as
$$
(d\bar{\psi}d\psi) \Rightarrow (d\bar{q}dq),
$$
but there appears additional term from Jacobian,
\begin{equation}
\mbox{Tr}\Big(\log (1-{1\over 2}D)\Big)=\mbox{Tr}\Big(-{D\over 2}
-{1\over 2} \Big({D\over 2}\Big)^2-\cdots\Big).
\label{Jacob}
\end{equation}
It is verified that at low-orders of $t$ the above factor does not
contribute.
However it is expected that nontrivial terms which depends on
the gauge fields will appear at sufficiently high-order of $t$.
This problem is currently under study and the results will
be reported in a future publication \cite{IN-QCD}.

\subsection{Positive $A$}

In the previous section we showed that for negative $A$
the system has desired properties concerning the chiral 
symmetry and we shall call this phase QCD phase.
In this section we shall briefly study the case of positive $A$.
For positive $A$, the $t$-expanded $D(m,n)$ is given as
\begin{equation}
D(m,n)=2\delta_{mn}+{C\over |A|}\sum\gamma_\mu\Gamma^-_\mu(m,n)
+O(t^2).
\label{DpA}
\end{equation}
Therefore there exists term like $\sum \bar{\psi}(m)\psi(m)$ in the action
besides the quark mass term.
One may think that this term breaks the extended chiral symmetry.
However this is not the case.
Actually from (\ref{extended2}),
\begin{eqnarray}
\delta\psi(m)&=&\epsilon\gamma_5\Big(-\delta_{mn}
-{C \over |A|}
\sum\gamma_\mu\Gamma^-_\mu(m,n)+\cdot\cdot\cdot\Big)\psi(n), \nonumber  \\
\delta\bar{\psi}(m)&=&\epsilon\bar{\psi}(m)\gamma_5,
\label{extended3}
\end{eqnarray}
and therefore $\psi$ and $\bar{\psi}$ have opposite 
``extended-chirality" with each other
(at leading order of $t$).
This fact suggests that the phase of positive $A$ is
different from that of negative $A$.\footnote{Strictly
speaking, we cannot deny the
possibility that these states are connected by a crossover
rather than a phase transition,
since our analysis cannot be applied for small $|A|$.
However existence of a phase transition between
them is plausible. See later discussion.}

Analysis of the effective action at the strong-coupling limit in the 
previous section can be applied also for the case of positive $A$,
and it is shown that condensation $\langle\bar{\psi}\psi\rangle$
has nonvanishing value.
However in the case of positive $A$,
\begin{eqnarray}
&\mbox{limit}&\Bigg[\langle\bar{\psi}\psi\rangle_{M_B}
+\langle\bar{\psi}\psi\rangle_{-M_B}\Bigg]\neq 0.  \nonumber  \\
&M_B \rightarrow 0&
\label{cond+}
\end{eqnarray}
Actually the effective potential for 
$\langle{\cal M}^\alpha_\beta\rangle=-v \delta^\alpha_\beta$
is given as
\begin{equation}
{V_{eff}\over NN_{sf}}=-2v +\log v+\cdot\cdot\cdot,
\nonumber
\end{equation}
and therefore $v={1\over 2}+O(t)$.

Condensation $\langle\bar{\psi}\psi\rangle$ does not mean the 
spontaneous breaking of the {\em extended} chiral symmetry.
Natural candidate for order parameter in the positive $A$ phase
is $\langle\bar{\psi}\gamma_\mu\psi\rangle$, for
\begin{equation}
\langle\delta(\bar{\psi}\gamma_\mu\gamma_5\psi)\rangle
=\langle\bar{\psi}\gamma_\mu\psi\rangle+O(t).
\end{equation}
However the strong-coupling analysis similar to that for negative $A$
shows $\langle\bar{\psi}\gamma_\mu\psi\rangle=0$.
Next one is the nearest-neighbor quark bilinear
$\langle\bar{\psi}\gamma_\mu U_\mu \psi\rangle$.
From the analysis at the strong-coupling limit (\ref{expectU}),
we can expect nonvanishing expectation value of the above
operators.
Moreover their values depend on the direction, i.e.,
\begin{equation}
\langle\bar{\psi}(m)\gamma_\mu U_\mu(m)\psi(m+\mu)\rangle
=-\langle\bar{\psi}(m+\mu)\gamma_\mu 
U^\dagger_\mu(m)\psi(m)\rangle\neq 0.
\label{NNcond}
\end{equation}
As
\begin{equation}
\bar{\psi}(1-{a \over 2}D)\psi(m)=-{C\over 2|A|}\bar{\psi}(m)
\sum\gamma_\mu\Gamma^-_\mu(m,n)\psi(n)+O(t^2),
\end{equation}
the condensation (\ref{NNcond}) means
\begin{equation}
\langle \bar{\psi}(1-{a \over 2}D)\psi\rangle\neq 0.
\label{OP2}
\end{equation}
Then from the criterion in \cite{chand} we can expect appearance 
of a massless particle at the channel $(\bar{\psi}\gamma_5\psi)$,
though meaning of the extended chiral symmetry is quite different from
the usual chiral symmetry in this phase.


\section{Discussion}

In this paper we study properties of 
the lattice QCD with the overlap fermions at strong coupling.
To this end, we slightly generalize the ordinary overlap fermion
by introducing the parameter $t$.
We expand the Dirac operator of the GO fermion in powers of $t$,
and then apply the standard techniques of the strong-coupling expansion.

It is important and urgent to examine validity and applicability
of the $t$-expansion, for a drawback of the overlap fermion is
its nonlocality.
By numerical calculation it is verified that for smooth configurations
of the gauge field the locality is satisfied \cite{smooth}.
On the other hand, we also expect that the locality is satisfied even 
at strong gauge coupling {\em after} integration over the gauge field
in certain parameter region of the GO fermion.
Results in the present paper support this expectation but more intensive
studies are required.
Tractable models in low dimensions might be useful.
Both numerical and analytical studies on them are needed in order
to argue the applicability of the $t$-expansion.
%
%
%
\begin{figure}[ht]
%
%
\begin{center}
\includegraphics[height=7.5cm]{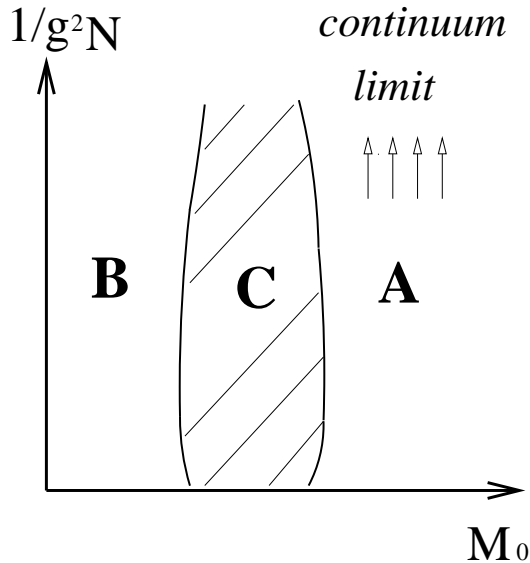}
  \caption{
    Schematic phase diagram of the lattice QCD with the overlap
    fermions in the $({1 \over g^2 N},M_0)$ plane.
    Phase A has desired properties of QCD.
    Extended chiral symmetry is spontaneously broken and 
    quasi-massless pions appear.
    On the other hand, the phase B is anomalous.
    Phase C in between is the nonlocal phase in which the long-distance
    terms give important effects. 
    The critical lines which separate phases A, B and C may have strong 
dependence on the gauge-coupling  constant $g^2$ though they are almost 
vertical in the figure.}
\label{phase}
\end{center}
\end{figure}

We find that there are (at least) two phases in the lattice QCD 
with the GO fermions at strong coupling.
One of them has desired properties of QCD.
Though this result is obtained by using the $t$-expansion,
we expect that it is correct even for the ordinary overlap fermion
system since expansion parameters are ${B\over A}$ and ${C\over A}$.
Therefore for sufficiently large $|A|$ our results are applicable.
In Fig.1, we show a possible phase diagram of the lattice QCD with 
the (generalized) overlap fermions.
The phase A has desired chiral properties of QCD and we call it 
QCD phase.
On the other hand, the phase B is anomalous as we explained in Sect.3.
The phase C in between cannot be studied by the present techniques,
for the $t$-expansion is not applicable.
There it is expected that nonlocal-long-distance terms cannot be 
neglected and they give substantially important effects on physical 
properties.
For example, the Goldstone theorem assumes the locality of the system
as is well-known.
Therefore in the phase C, which we call nonlocal phase, massless
pions might not appear even if the extended chiral symmetry is 
spontaneously broken.
It is possible that the nonlocal phase C does not exist in certain
parameter region of $(r,t)$.

Detailed study on the QCD phase will be reported in a 
forthcoming paper \cite{IN-QCD}.
Especially, it is interesting to see how the U(1) problem 
is solved and how its relates with the noninvariance of the
fermion path-integral measure under flavor-singlet extended
chiral transformation \cite{Lus}.
In the framework of the $t$-expansion, the Jacobian 
$\mbox{Tr}(\gamma_5D)$ is easily evaluated at low-orders of $t$,
and it is verified that term corresponding to the anomaly appears. 
However its coefficient is not constant but depends on $t$.
Another interesting problem is the strong-coupling chiral 
gauge theory \cite{chiral}.
This system might be studied by using the techniques in this paper.

\vskip 0.5cm
\begin{bf}
\noindent
\large{Note added}
\end{bf}
 
After submitting this paper, there appeared an interesting
paper\cite{hip} which discusses approximate solutions of the 
G--W relation which are useful for numerical simulations.

\newpage


\begin{thebibliography}{99}%

\bibitem{NN}R.Narayanan and H.Neuberger, Nucl.Phys.
B412(1994)574;\\
Nucl.Phys.B443(1995)305.
%
\bibitem{Ne}H.Neuberger, Phys.Lett.B417(1998)141. 
%
\bibitem{GW}P.H. Ginsparg and K.G. Wilson, 
Phys. Rev. D25(1982) 2649.
%
\bibitem{Lus}M.L\"uscher,
Phys.Lett.B428(1998)342.
%
\bibitem{BG}E.Brezin and D.J.Gross,
Phys.Lett.97B(1980)120.
%
\bibitem{I}I.Ichinose, 
Nucl.Phys.B249(1985)715.
%
\bibitem{KS}N.Kawamoto and J.Smit,
Nucl.Phys.B192(1981)100.
%
\bibitem{chand}S.Chandrasekharan,
Phys.Rev.D60(1999)074503.
%
\bibitem{IN2}I.Ichinose and K.Nagao,
hep-lat/9909035.
%
\bibitem{IN-QCD}I.Ichinose and K.Nagao,
hep-lat/0001030.
%
\bibitem{smooth}P.Hernandez, K.Jansen and M.L\"uscher,
Nucl.Phys.B552(1999)363.
%
\bibitem{chiral}M.L\"uscher,
Nucl.Phys.B538(1999)515; Nucl.Phys.B549(1999)298; \\
hep-lat/9904009.
%
\bibitem{hip}C.Gattringer and I.Hip,
hep-lat/0002002.

 %
\end{thebibliography}
\end{document}